\documentclass[aps,prx,reprint]{revtex4-2}
\usepackage[utf8]{inputenc}
\usepackage[T1]{fontenc}
\usepackage{bbold}
\usepackage{amsmath,mathtools}
\usepackage{times}
\usepackage{mathrsfs}

\usepackage{calc} 
\usepackage{graphicx}
\usepackage{tabularx}
\usepackage{colortbl}
\usepackage[citecolor=blue, urlcolor=blue, linkcolor=blue, colorlinks=true]{hyperref}

\graphicspath{{figures/}}

\newcommand*\dagg{^{\dagger}}	

\newcommand*{\ket}[1]{|#1\rangle}
\newcommand*{\bra}[1]{\langle#1|}

\newcommand{\eps}{\varepsilon}

\newcommand{\Ry}{\mathrm{Ry}}
\newcommand{\up}{\uparrow}
\newcommand{\down}{\downarrow}

\usepackage[capitalize]{cleveref}
\crefformat{section}{Sec.~#2#1#3}

\begin{document}
\title{Few-body analogue quantum simulation with Rydberg-dressed atoms in optical lattices}
\author{Daniel Malz}
\altaffiliation{Current address: Department of Physics, Technische Universität München, James-Franck-Straße 1, 85748 Garching, Germany}
\affiliation{Max Planck Institute for Quantum Optics, Hans-Kopfermann-Straße 1, D-85748 Garching, Germany}
\affiliation{Munich Center for Quantum Science and Technology, Schellingstraße 4, D-80799 München, Germany}
\author{J.~Ignacio Cirac}
\affiliation{Max Planck Institute for Quantum Optics, Hans-Kopfermann-Straße 1, D-85748 Garching, Germany}
\affiliation{Munich Center for Quantum Science and Technology, Schellingstraße 4, D-80799 München, Germany}

\begin{abstract}
  Most experiments with ultracold atoms in optical lattices have contact interactions, and therefore operate at high densities of around one atom per site to observe the effect of strong interactions.
  Strong ranged interactions can be generated via Rydberg dressing, which opens the path to explore the physics of few interacting particles.
  Rather than the unit cells of a crystal, the sites of the optical lattice can now be interpreted as discretized space.
  This allows studying completely new types of problems in a familiar architecture.
  We investigate the possibility of realizing problems akin to those found in quantum chemistry, although with a different scaling law in the interactions.
  Through numerical simulation, we show that simple pseudo-atoms and -molecules could be prepared with high fidelity in state-of-the-art experiments.
\end{abstract}

\maketitle

\section{Introduction}
Analogue quantum simulation with ultracold atoms trapped in optical lattices has proven to be a highly successful tool to realize and study interacting many-body systems in the laboratory~\cite{Bloch2008}. In a typical setup, neutral atoms are loaded with an average density of one per lattice site, and interact via contact interactions (Hubbard-$U$), realizing a Fermi- or Bose-Hubbard model~\cite{Jaksch1998}.
Site-resolved detection in a quantum gas microscope~\cite{Bakr2009,Haller2015} and the ability to apply a site-resolved potential~\cite{Choi2016} add to a rich toolbox to study many-body systems~\cite{Gross2017}.
The focus on densities of around 1 has two primary reasons.
First, because this is the regime corresponding to strongly correlated electrons in condensed matter.
Second, relatedly, because the existing experiments are usually limited to contact interactions, which means that high densities are required to obtain strong interaction effects.

Here, we would like to explore quantum simulation of few-particle systems, \emph{i.e.}, the regime of low densities, which in turn requires ranged interactions.
This is motivated by the ubiquity of long-range forces at the nanoscale~\cite{French2010}, which makes it important for fields such as quantum chemistry~\cite{Bauer2020}.
Rydberg states are a powerful way to obtain strong ranged interactions between neutral atoms~\cite{Jaksch2000,Lukin2001}, which can also be used for quantum simulation~\cite{Weimer2010,Browaeys2020}.
These interactions decay as $r^{-6}$ or $r^{-3}$~\cite{Pupillo2010}.
Since the Rydberg interaction is much stronger than hopping, we require Rydberg dressing~\cite{Bouchoule2002,Balewski2014,Zeiher2016} to make the interaction tunable and comparable to hopping rates.
This setting, of atoms hopping in a two-dimensional lattice and repelling each other via ranged Rydberg interactions, is reminiscent of electrons in two-dimensional discretized space, albeit with the repulsive force following a different power law than typical Coulomb interactions.
We note that there exist experiments that have successfully explored few-particle physics but with a different focus and context, notably experiments with weakly interacting fermions in the continuum~\cite{Bayha2020,Holten2021,Holten2022}, or low densities of Rydberg excitations~\cite{Zeiher2016,Bernien2017,Hollerith2019}.

\begin{figure}[t]
  \centering
  \includegraphics[width=\linewidth]{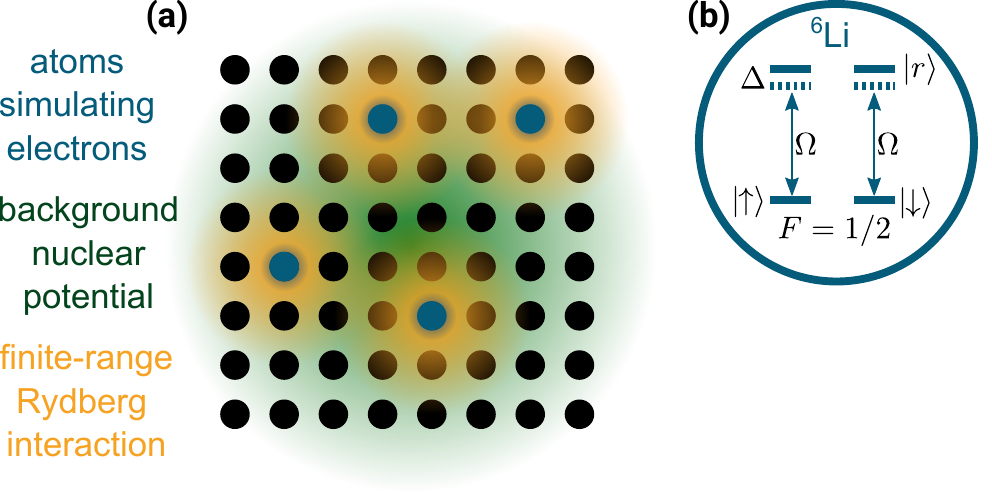}
  \caption{\textbf{Proposed setup.}
  	Fermionic atoms (blue) with two hyperfine ground states (corresponding to spin-up and spin-down) are trapped in an optical lattice (black). The strength of the trapping laser controls the hopping strength $J$. The atoms, which play the role of electrons, experience a site-dependent Stark shift that serves to emulate a nuclear potential (green).
  	Repulsive interactions between the atoms are induced through Rydberg dressing (yellow).
  	Right: example level structure. Rydberg dressing is achieved through a laser of strength $\Omega$ that is detuned by $\Delta$ from the transition from the $F=1/2$ hyperfine ground states to a Rydberg state in the $p$ shell.
  }
  \label{fig:setup}
\end{figure}

Analogue quantum chemistry with ultracold atoms has been considered before.
The possibility of simulating molecular orbitals was raised in Ref.~\cite{Luhmann2015}.
Full analogue simulation of quantum chemistry with cold atoms was proposed in Refs~\cite{Arguello-Luengo2019,Arguello-Luengo2020}.
Although this shows that such an approach is feasible in principle, it is difficult to make it practical with existing technology. 
Apart from creating three-dimensional potentials and three-dimensional site-resolved detection, another key difficulty in that proposal is the repulsive Coulomb interaction that decay as $1/r$, which require a cavity and another atomic array to implement.
This motivates the search for other platforms (see also Ref.~\cite{Knorzer2022}).
In the setup proposed here, with Rydberg dressing and in two dimensions, one clearly cannot achieve quantitative agreement with quantum chemistry, but as we show, it is nevertheless possible to simulate what we call ``pseudo quantum chemistry,'' which is conceptually and qualitatively much like real quantum chemistry.
More generally, long-range interactions pose considerable challenges to existing numerical methods, which makes quantum simulation of them a particularly worthwhile target. 

As we demonstrate with numerics below, in the proposed setup one can reproduce central phenomena of quantum chemistry qualitatively, and thus study the physical effects that arise in such systems experimentally.
Moreover, the setup is simple, and the relevant tools have already been showcased experimentally~\cite{Guardado-Sanchez2021}.
Similar to the dense regime (e.g.\ Ref.~\cite{Madhusudhana2021}), this analogue quantum simulator can also be used to benchmark classical methods and push the state of the art in classical numerical techniques for interacting few-body problems.
Our proposal not only enables pseudo quantum chemistry simulations, but it also opens new possibilities in ultracold atom experiments.

\section{Setup}
Our proposal requires fermionic atoms hopping in an optical lattice, with two long-time coherent hyperfine levels that can be dressed to the same Rydberg state (see illustration in \cref{fig:setup}).
We furthermore require a potential with ideally close to single-site resolution, which can be engineered using AC Stark shifts.
Shaping of the laser to achieve a tunable potential can be done for example using a digital mirror device and the objective of a quantum gas microscope~\cite{Choi2016}, or through spatial light modulators~\cite{Gaunt2012,Sturm2017}.
For preparation and readout, a quantum gas microscope is essential, because we consider the initial state of the experiments to be atoms localized at known lattice sites.

The Hamiltonian of the system is given through
\begin{equation}
  H = H_{\mathrm{kin}} + H_{\mathrm{pot}} + H_{\mathrm{int}}.
  \label{eq:H}
\end{equation}
The noninteracting Hamiltonian, $H_0=H_{\mathrm{kin}}+H_{\mathrm{pot}}$, comprises the hopping in the lattice and a background potential
\begin{equation}
  H_0 = -J\sum_{\langle \vec i \vec j\rangle,\sigma}c_{\vec i\sigma}\dagg c_{\vec j\sigma}
  -\sum_{\vec i, \sigma}V_{\mathrm{pot},\sigma}(\vec i)c_{\vec i\sigma}\dagg c_{\vec i\sigma},
  \label{eq:H0}
\end{equation}
with hopping rate $J$, and where $c_{\vec i,\sigma}\dagg$ creates a fermion with spin $\sigma\in\{\up,\down\}$, encoded by two hyperfine states of the fermion, on the lattice site $\vec i$.
Note that we allow the potential to be spin-dependent in general.

We propose to use Rydberg dressing~\cite{Bouchoule2002} to engineer repulsive interactions.
Using a laser of Rabi frequency $\Omega$ and detuning $\Delta$ from the transition to a Rydberg state $\ket r$, the ground state of the atom is dressed to $\ket g_{\mathrm{dr}}\approx\ket g+\beta\ket r$, with $\beta=\Omega/2\Delta$, where $\Omega\ll\Delta$ (see \cref{fig:setup}).
At large interatomic distances, this yields a repulsive interaction $V(\vec r)\approx \beta^4 V_{\mathrm{ryd}}(\vec r)$, where $\vec r$ is the distance between two dressed atoms and $V_{\mathrm{ryd}}$ is the interaction potential for two Rydberg atoms. 
At small interatomic distances (of the order of a lattice distance), the potential saturates to $\Omega^4/(2|\Delta|)^3$~\cite{Zeiher2016}.
The presence of this soft-core potential does not qualitatively change the physics that we explore here, but will lead to quantitative differences.
For simplicity, in the following we consider an interatomic interaction of the form
\begin{equation}
  H_{\mathrm{int}} = V_{\mathrm{int}}\sum_{\vec i\neq \vec j,\sigma,\rho}\frac{1}{|\vec i-\vec j|^\alpha}n_{\vec i\sigma}n_{\vec j\rho}
  +2V_{\mathrm{int}}\sum_{\vec i}n_{\vec i\up}n_{\vec i\down}
  \label{eq:Hint}
\end{equation}
where $n_{\vec i\sigma}=c_{\vec i\sigma}\dagg c_{\vec i\sigma}$ is the number operator, and $\alpha$ controls the decay of the interaction.
We will consider two different mechanisms to achieve isotropic repulsive interactions: first, we consider standard Rydberg dressing, which results in interactions that decay with $\alpha=6$, and second, we consider Rydberg dressing in the presence of a strong electric field, which yields interactions that decay as $\alpha=3$~\cite{Jaksch2000,Pupillo2010}.

\subsection{Motivation: pseudo quantum chemistry}
One major motivation to study problems of the type \cref{eq:H} with low particle density is its similarity to quantum chemistry.
One of the most important problems in quantum chemistry is to find the ground state of $N_e$ electrons in an attractive potential provided by $N_{\mathrm{nuc}}$ nuclei, where according to the Born-Oppenheimer approximation, the nuclei are considered immobile~\cite{Aspuru-Guzik2005}.
The quantum chemistry Hamiltonian is then composed of the kinetic energy term, the nuclear potential, and the repulsive particle--particle interactions.
With this perspective, $H_{\mathrm{kin}}$ can be interpreted as the motion of electrons in two-dimensional discretized space, $H_{\mathrm{int}}$ as the Coulomb repulsion (albeit with a different power law decay), and the background potential can be chosen as
\begin{equation}
  H_{\mathrm{pot}}=H_{\mathrm{nuc}}=-\sum_{\vec i}\sum_nV_{\mathrm{nuc}}^{(n)}(\vec i)c_{\vec i}\dagg c_{\vec i},
  \label{eq:Hnuc}
\end{equation}
where $V_{\mathrm{nuc}}^{(n)}(\vec r)=JZ_n/(a_0|\vec r_n-\vec r|)$ is the attractive potential produced by the $n^{\mathrm{th}}$ nucleus.
We have assumed that there are $N_{\mathrm{nuc}}$ nuclei, placed at positions $\vec r_n$ and with charge $Z_n$.
The strength of the nuclear potential relative to the hopping is controlled by $a_0$, which sets the Bohr radius in units of the lattice constant, and thus the Rydberg constant $\Ry=J/a_0^2$.
We note that a $r^{-1}$ nuclear potential is not strictly required to see phenomena akin to quantum chemistry and qualitatively similar physics can be observed in a variety of nuclear potential shapes~\footnote{
  The two-dimensional geometry allows essentially arbitrary on-site potential to be applied from the transverse direction.
  This leaves considerable freedom in choosing a nuclear potential. 
  Natural choices are $V_{\mathrm{nuc}}(r)\propto \log(r)$, which is the elector-magnetic potential in flatland consistent with Gauss' law, or $V_{\mathrm{nuc}}(r)\propto r^{-\alpha}$ for integer $\alpha$.

  The flatland potential $V_{\mathrm{nuc}}(r)\propto \log(r)$ has been investigated in Ref.~\cite{Atabek1974}.
  This potential obeys Gauss' law in two dimensions, which means that if the ``electrons'' could be made to interact according to the same potential, then at large distances, a two-dimensional atom would appear to have a charge equal to the coordination number minus the number of bound electrons.
  However, since Rydberg dressing can only produce interactions that scale as $r^{-3}$ or $r^{-6}$, we have to sacrifice Gauss' law.

  It therefore makes sense to go to a more localized nuclear potential as well, which improves convergence with system size and makes it easier to have strong interactions.
  It is tempting to choose $V_{\mathrm{nuc}}(r)\propto r^{-\alpha}$ with $\alpha=3$ or $\alpha=6$ to match it to the electronic interaction potential. 
  However, an attractive potential with $\alpha\geq2$ leads to a Hamiltonian that is not lower bounded.
  This can easily be seen by inserting a trial wavefunction $\psi(r)= \exp(-r^2/(2\sigma^2))/\sqrt{\pi\sigma^2}$ into the Hamiltonian,
  which yields
  \begin{equation}
  	\langle \psi|H|\psi\rangle =\int_0^\infty\frac{2\pi r\,dr}{\pi\sigma^2} e^{-r^2/\sigma^2}
  	\left( -r^{-\alpha}+\frac{2}{\sigma^2}-\frac{r^2}{\sigma^4} \right).
  	\label{eq:bound_alpha}
  \end{equation}
  While the kinetic energy is finite, the integral giving the potential energy only converges for $\alpha<2$.

  Thus, as a compromise, we propose to use $V_{\mathrm{nuc}}(r)=-V_0/r$. The system can then be interpreted as a standard atom, but with the electron confined to move in a plane. In this scenario we can identify $V_0=e^2/(4\pi\eps_0)$.
  This potential has the advantage that the single-particle wavefunctions can be obtained analytically through standard procedure~\cite{Zaslow1967}.
}
In the present choice, the nuclear potential can be interpreted as being generated by a Coulomb force in three dimensions, but with the electron confined to move in a plane. In this scenario we can identify $V_0=e^2/(4\pi\eps_0)$, and one can find the atomic orbitals analytically in the continuum~\cite{Zaslow1967}, with principal energies $E_n=-\Ry/(n-1/2)^2$.
A more experimentally friendly choice could be a Gaussian potential produced by a tightly focused laser beam.

We note that the Hamiltonian $H=H_0+H_{\mathrm{int}}$ does not include spin-orbit coupling; it does not couple spin-up to spin-down particles.
Nevertheless, spin is explicitly included and has important consequences, as we explore below.
Furthermore, $H_{\mathrm{pot}}$ could include a linear gradient, which would model an electric field (a potential gradient coupling independent of spin) or a gradient with opposite signs for spin-up and spin-down particles, which corresponds to an applied magnetic field.

In the following, we focus on the pseudo quantum chemistry regime for concreteness, and so we take the background potential to be the nuclear potential produced by some distribution of nuclei.
In our simulations below we make the choice $V_{\mathrm{int}}=Ja_0^{\alpha-2}/\alpha$, which is chosen such that the nuclear attraction is equal to the electronic repulsion at $r=a_0$, and we take $\alpha=6$.

\subsection{Preparation and measurement}
To find the ground state energy requires preparing the ground state with high fidelity, and subsequently measuring its energy. We propose to prepare the ground state adiabatically.
There is substantial freedom in designing adiabatic paths. Here we propose to take the simplest choice: first prepare the non-interacting electronic state, and then adiabatically turn on interactions. This has the advantage that during the majority of the time, the interaction (and thus Rydberg decay) is turned off. 
Starting with exactly $N_e$ atoms, localized at known sites in the optical lattice, and no hopping, first
prepare the desired spin state, which can be done using the single-site access afforded by a quantum-gas microscope~\cite{Gross2021}.
Second, tune the noninteracting Hamiltonian $H_0$ adiabatically to its final form.
This requires a Hamiltonian path that takes the $N_e$ electrons on their initial sites (which are eigenstates when hopping is turned off) to the desired molecular orbitals.
In a final step, turn on interactions adiabatically to reach the interacting ground state.

To understand how the adiabatic preparation time generally depends on the Bohr radius $a_0$, note that the gaps between the atomic orbitals are proportional to the Rydberg constant $\mathrm{Ry}=J/a_0^2$, which intuitively corresponds to the rate at which an atom can explore an area equal to an orbital of radius $a_0$.
Since adiabatic preparation generally scales with the inverse of the gap squared~\cite{Albash2018}, we expect that the adiabatic preparation time scales as $\Ry^{-1}\propto a_0^4$.

To measure the preparation fidelity, perhaps the simplest approach is to reverse the adiabatic preparation and then to measure the particle distribution. To a good approximation, the probability for the atoms to return to their initial positions is the square of the fidelity of the ground state preparation. 

Rather than measuring the occupancy of the ground state, it is also possible to measure the energy of the system directly.
This requires measuring the kinetic and potential energy of the atoms.
The potential energy can be computed straightforwardly from the atomic configuration in real space, and thus measuring it requires taking a snapshot of the system with a quantum gas microscope. 
The kinetic energy requires determining the momentum of each particle, which may be done using the time-of-flight detection, provided all atoms can be caught.
Such single-atom momentum measurements have already been realized~\cite{Cayla2018,Carcy2019,Holten2021}.

Finally, below we also discuss the potential to do spectroscopy directly on the system.

\subsection{Experimental imperfections}
Rydberg dressing leads to the main source of decoherence for this protocol, because the admixed Rydberg state may decay~\cite{Balewski2014}.
Given a dressing $\ket g_{\mathrm{dr}}=\ket g +\beta \ket r$, the interaction at large distances scales with $\beta^4$, whereas the probability for a Rydberg decay scales as $\beta^2$. Since we require $\beta\ll1$ to make the hopping and Rydberg interaction rates comparable, the ratio of interaction strength to decay rate is multiplied by a small factor $\beta^2$ relative to equivalent ratio of the Rydberg state.
If an atom is excited to the Rydberg state it is typically lost rapidly, as is observed in experiment~\cite{Zeiher2016,Guardado-Sanchez2021}, which is why in the experiments one can postselect onto runs in which the atom number remains unchanged to remove this effect.
However, this still leads to a limitation, as the success probability of the experiment decays exponentially with time, which means that Rydberg decay effectively limits the accessible time scales.
Thus, the main focus of our numerical simulations is to assess whether the adiabatic preparation can succeed within the time available.

\begin{figure}[t]
  \centering
  \includegraphics[width=\linewidth]{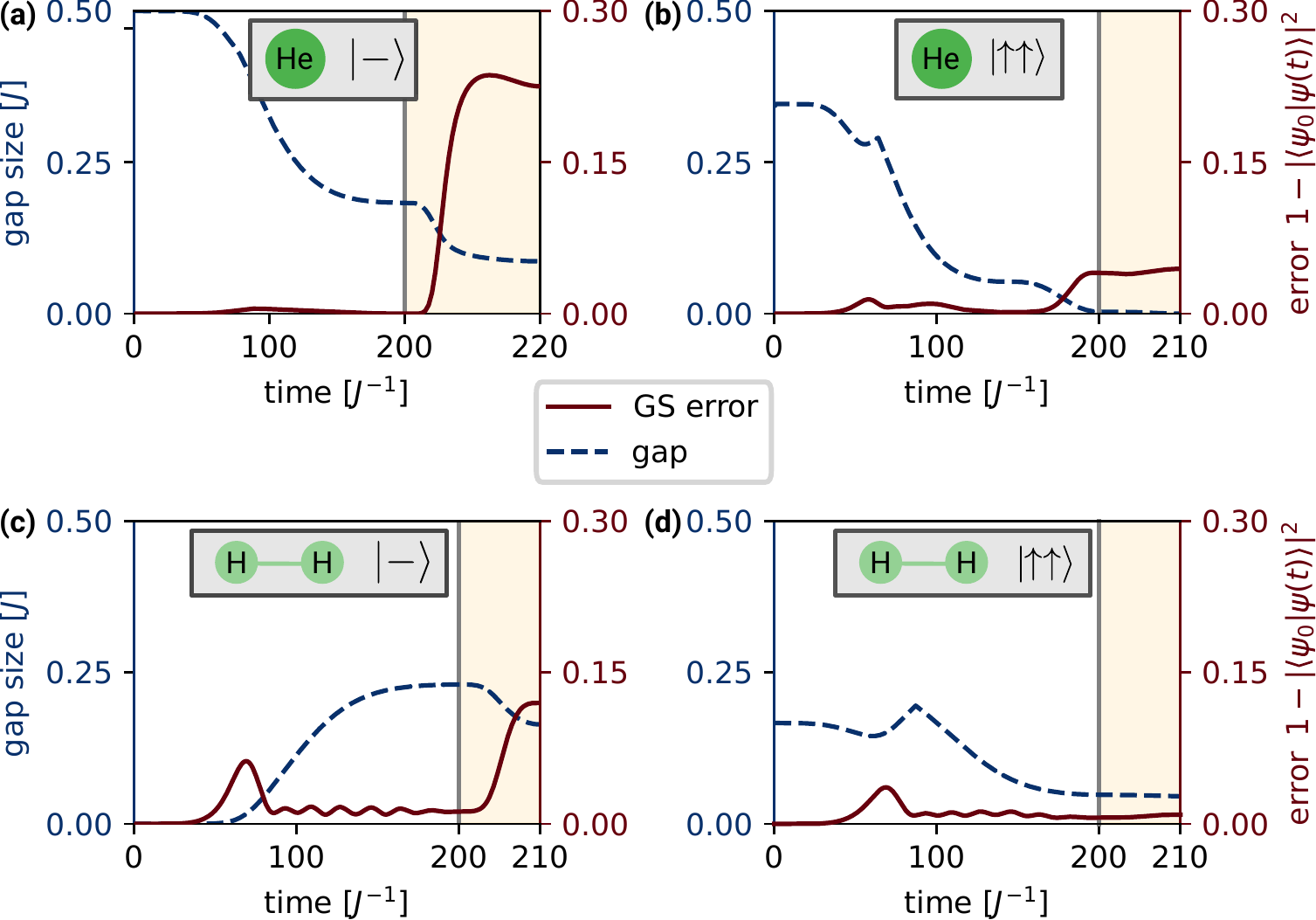}
  \caption{\textbf{Adiabatic preparation of pseudo He and H$_2$.}
  	We illustrate the adiabatic preparation by example of He and H$_2$.
  	At $t=0$ we start with two fermions in a spin singlet $\ket{-}$ (left, (a,c))
  	or a triplet $\ket{\!\!\up\up}$ (right, (b,d)).
  	The adiabatic preparation proceeds by preparing the noninteracting ground state up to $t=200J^{-1}$ (vertical grey line) and subsequently turning on the interaction adiabatically (yellow shaded area).
  	For details on the preparation, we refer to the main text.
  	The system parameters used to simulate the preparation of He were $a_0=4$, and central charge $Z=2$, with the nucleus placed in the centre of a $21\times21$ lattice.
  	For H$_2$, we separated two nuclei with $Z=1,\,a_0=2$ by three lattice sites, and centred them in a $24\times21$ lattice.
  	As interactions, we always take $\alpha=6$ and $V_{\mathrm{int}}=J a_0^{\alpha-2}/\alpha$.
  	If the fermions are in a spin singlet, their spatial wavefunction is bosonic, such that they occupy the same spatial orbital during the non-interacting preparation, such that interactions have a large impact on the state, in contrast to the spin triplet case.
  	Note that in panel (a), for bosons in He, the interaction time is twice as long as in the other panels.
  	The shown errors for the wavefunction lead to a relative errors in the ground state energy prediction of about 5\% (a), 35\% (b), 0.6\% (c), 7\% (d).
  	Note that in the figure, interacting region $Jt>200$ is stretched for presentation reasons.
  }
  \label{fig:prep}
\end{figure}

In Ref.~\cite{Guardado-Sanchez2021}, Rydberg dressing resulted in a lifetime of about $\tau_{\mathrm{eff}}=1$\,ms at an interaction strength comparable to the hopping of $J=2\pi\times 1.7$\,kHz, leading to a figure of merit of about $J\tau_\mathrm{eff}\approx 10$.
We note that in the same experiment, the widely observed avalanche loss~\cite{Balewski2014,Goldschmidt2016,Zeiher2016,Festa2022} was found to be almost negligible at small densities, so we will neglect this effect in the following.
Below we show that even in a time of about $10J^{-1}$, the interacting ground state of simple systems can be prepared with fairly high fidelity (\emph{cf}.\ \cref{fig:prep}), placing the experiments proposed here into experimental reach. 

Other experimental imperfections include disorder in the nuclear potential, which arises because the optical lattice spacing is comparable to or below the diffraction limit, which makes it difficult to accurately implement the potential, as well as measurement and preparation errors.
Errors in the initial state preparation will directly translate to errors in the measured quantities here, but in principle this can be controlled by imaging the atoms before the start of the experiment.
Finally, like in all ultracold atom experiments, collisions with background gas and heating through laser fluctuations must be controlled sufficiently as to not influence the experiment during the preparation and simulation time.

\section{Results}
In this section, we numerically simulate a few potential near-term experiments using exact diagonalization.
Exact classical simulation of the system is expensive, and so we are limited to small system sizes and few atoms.

\subsection{Adiabatic preparation of pseudo atoms and molecules}\label{sec:noninteracting_adiabatic}
We explore adiabatic preparation through the examples of two fermions in one or two potential sites, analogous to helium and molecular hydrogen.
Since the most stringent limitation on current experiments is likely the Rydberg decay, which limits the total interaction time, we focus on assessing how well preparation works for realistic interaction times.
We study both the case in which the two particles are in a spin singlet (corresponding to the actual ground state), in which case they behave like bosons, and the case in which they are in a spin triplet (e.g., both in the same hyperfine ground state), where they behave like fermions.

The simplest case is bosonic helium (\cref{fig:prep}a), because both particles occupy the same orbital.
In our simulations, we assume that there is a single nucleus with $Z=2$ in the middle of a $(2p+1)\times(2p+1)$ lattice and place both particles on the same site. In all simulations here, the padding $p=10$.
With the nuclear potential fixed, we take $J(t)=J\sin^4(\pi t/(2T))$ with $T=200J^{-1}$, which prepares both particles in the 1s orbital.
Afterwards, we turn on interactions in the same way, $V_{\mathrm{int}}(t)=V_{\mathrm{int}}\sin^4(\pi t/(2T_{\mathrm{int}}))$, with $T_{\mathrm{int}}=10$, while keeping $H_0$ fixed.

Fermionic helium (\cref{fig:prep}b) is somewhat more complicated as the noninteracting part requires to simultaneously prepare one particle in the 1s orbitals and one in one of the first excited orbitals. Since the system is finite, the 2s orbital is slightly higher in energy than the 2p orbitals.
We numerically break the degeneracy between the two 2p orbitals by adding a very weak attractive potential $V_{2p}(x)\propto|\psi_{2p}(x)|$, about a hundred times weaker than the nuclear potential.
We place one particle on the nucleus site, and a second particle three sites away. We then add another auxiliary nuclear potential at that site, at 90\% strength of the original one, such that this site has the second lowest potential.
In a two-step process, we first turn on hopping while keeping this potential configuration fixed (same $J(t)$ as above, but with $T=140J^{-1}$), and then adiabatically turn off the auxiliary nuclear potential, keeping still the weak potential $V_{2p}$.
In the last step, we turn on interactions adiabatically while turning off $V_{2p}$.

The preparation of molecular hydrogen (\cref{fig:prep}c and d) is conceptually more straightforward.
We always start with one particle on each of the two nuclei sites, turn on hopping adiabatically, and then interactions.

Already at this stage, particle statistics play a very important role. If we initialize the system in a spin singlet, the spatial wavefunction is symmetric (bosonic) and both particles occupy the same orbital. Turning on repulsive interactions thus has a large effect. 
If we instead initialize the system in a symmetric spin triplet state, the particles behave like fermions and are already well separated on average, and thus additional repulsive interactions hardly change the state. 
Most importantly, the wavefunctions can be prepared fairly accurately in most cases even with short interaction times, and the resulting energies have small error.

\subsection{Example two-dimensional chemistry experiment: molecular hydrogen}
\begin{figure}[t]
  \centering
  \includegraphics[width=\linewidth]{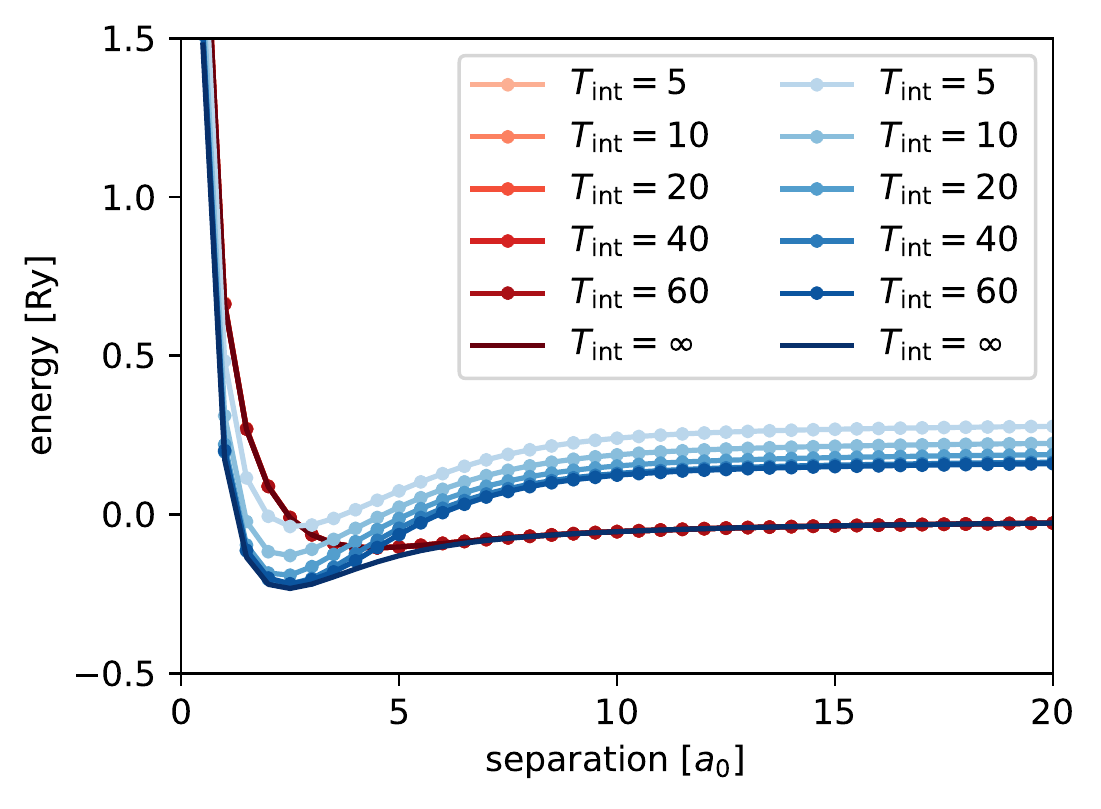}
  \caption{
  	\textbf{Simulated experimental determination of binding energy and bond length of pseudo-$H_2$.}
  	We simulate the adiabatic preparation of the ground states of pseudo-$H_2$ for a range of nucleus separations, and for either bosonic (blue) or fermionic particles (red). We plot the predicted ground state energy minus twice the energy of $H$ (corresponding to free hydrogen atoms).
  	We assume that the noninteracting wavefunction has been prepared error-free, and just study the error introduced due to adiabatically turning on interactions.
  	In a symmetric spin triplet state (red), the atoms have an antisymmetric spatial wavefunction, such that they occupy different orbitals, and thus the interaction hardly has any effect (the curves all lie on top of each other).
  	The molecule is very weakly bound in this case (hardly visible).
  	In the actual ground state of $H_2$, the electrons are in a spin singlet, and the optimal bond length is clearly visible, even at very short preparation times.
  	At large separations, the adiabatic preparation increasingly fails to produce the correct energy, as overlap of the single-particle wavefunctions decays exponentially with distance.
  	Parameters: $a_0=2$, padding $p=10$, $V_{\mathrm{int}}=a_0^{\alpha-2}/\alpha$.
  }
  \label{fig:H2_distance}
\end{figure}

With the ability to prepare the ground state of molecular hydrogen, we now explore how well bond length and ground state energy can be found with limited interaction time.
To this end, we simulate the adiabatic preparation for a range of nucleus separations and interaction times $T_{\mathrm{int}}$. 
The results are shown in \cref{fig:H2_distance}.

\begin{figure*}[t]
  \centering
  \includegraphics[width=\linewidth]{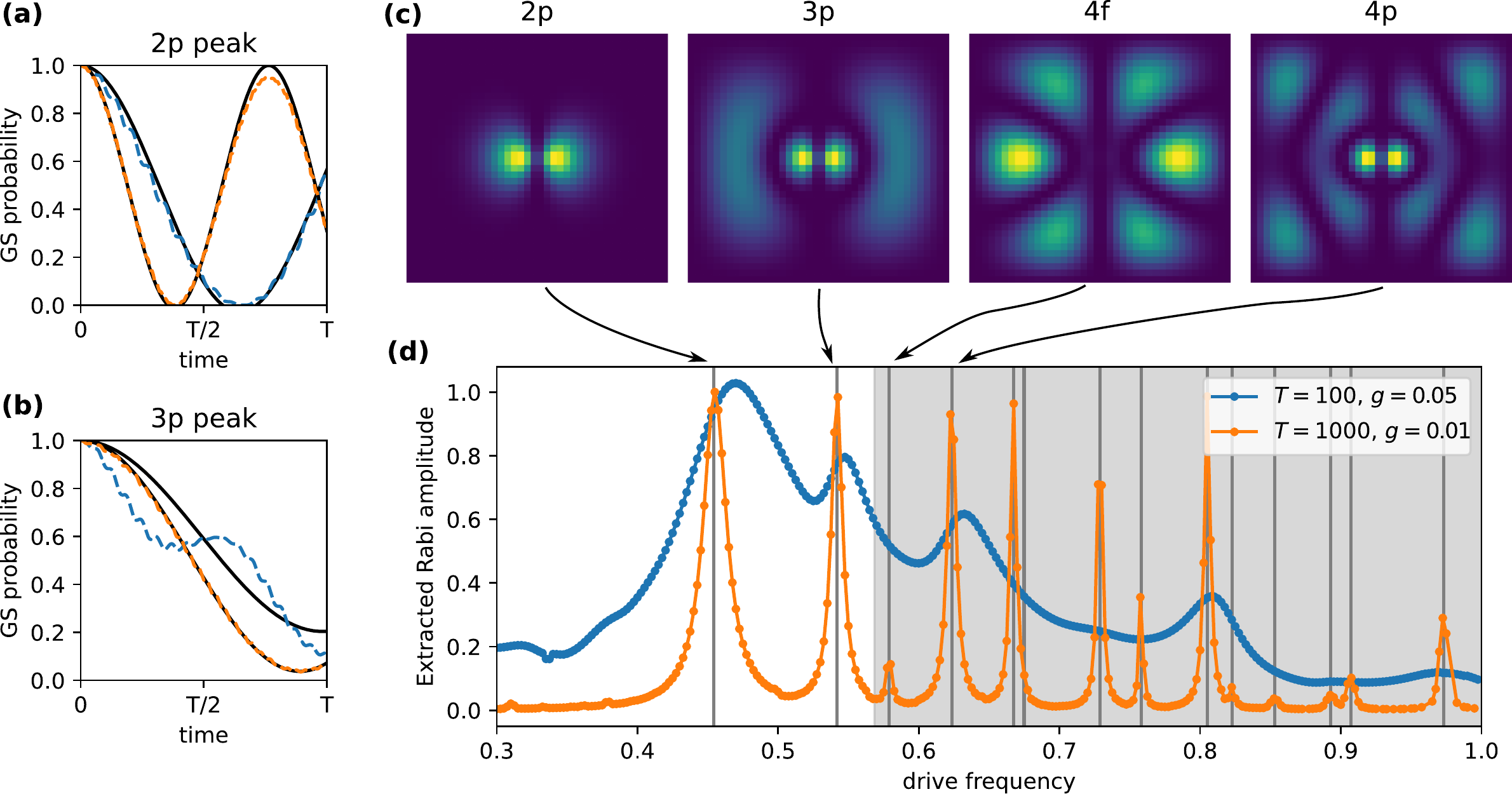}
  \caption{\textbf{Spectroscopy of hydrogen.}
  	Starting from a particle in the 1s orbital, an oscillating linear field is applied, corresponding to coherent driving in the dipole approximation.
  	The resulting dynamics is monitored by measuring the ground state probability as a function of time.
  	We simulate this process on a $40\times38$ lattice (to break symmetry for convenience) for a range of drive frequencies from $\omega=0$ to $\omega=J$ once for a driving strength $g=0.05$ and a total time of $T=100/J$ (blue) and once for a driving strength $g=0.01$ and a total time $T=1000/J$ (orange).
  	At each frequency, we fit $1-A\sin^2(\Omega t)$ to the numerically calculated ground state probability $|c_0(t)|^2$.
  	(a) Time trace of the first resonance, which corresponds to the 1s--2p transition. We plot the numerically calculated ground state probability (dashed coloured) and the fit (solid black). Note that the $x$ axis is scaled to the total time, which is $100/J$ (blue) and $1000/J$ (orange).
  	(b) The same plot for the 1s--3p transition. The blue curve markedly deviates from a pure sine function, as the effective coupling is fairly strong compared to the detuning to other states.
  	(c) Orbital wavefunctions of the excited states corresponding to the spectroscopy peaks.
  	(d) A plot of the fitted Rabi amplitude $A$ as a function of drive frequency. Weaker coupling requires longer times, but leads to a better resolution of the peaks. Note that the 1s--4f transition is forbidden, but since the system does not obey rotation symmetry exactly there exists a small nonzero matrix element.
  	The grey shaded area corresponds to transitions to states with an energy greater than $-4J$ (which in the infinite system would correspond to an unbound electron). The energy of these states is dominated by finite size effects. 
  }
  \label{fig:spectroscopy}
\end{figure*}

Our results indicate that even in a modest interaction time such as $T_{\mathrm{int}}=10J^{-1}$, one can already clearly see the molecular bound state and obtain an accurate result for the bond length.
Again, we also observe a clearly observable difference between fermionic and bosonic particles.
The bosonic case is much qualitatively much like one would expect, with a clearly defined molecular potential, even for short interaction times.
A short to intermediate distances, quantitative agreement with the true value is reached at still modest times between $20J^{-1}$ to $40J^{-1}$.
Interestingly, there evidently also is is a limitation of our chosen adiabatic path, in that it fails to give the right ground state at large separations. 
The reason is that in the initial noninteracting ground state, both particles occupy a symmetric orbital across both nuclei, which yields a 50\% chance of finding both particles on the same nucleus. 
Clearly, the true ground state at large separations should be two isolated hydrogen atoms, with zero probability of finding both particles at the same nucleus.
However, at large distances there is a potential barrier for particles to move between the nuclei, leading to exponentially small matrix elements between the states, which makes the adiabatic preparation fail.

In contrast, the molecular potential for fermionic particles hardly exists, to the extend that it is not properly visible on the plot.
In this case, there is no issue with obtaining the ground state adiabatically also at large separations, because the noninteracting ground state is already correct in that case.

\subsection{Spectroscopy}
In real atoms, coupling to the electromagnetic field allows excited electrons to relax to the ground state and coherent driving yields Rabi oscillations. In the dipole approximation, the atom is taken to be small as compared to wavelength of the emitted light.
In the present context, we can model coherent driving by applying a oscillating linear background potential, which allows us to observe dipole transitions.
In this section we focus on the noninteracting problem for simplicity. 
Experimentally, the interacting case works just as well as the non-interacting case, but the times required to perform spectroscopy demand about one order of magnitude improvements on the previously reported figure of merit of Rydberg dressing time over Rydberg decay time.
Spectroscopy through modulation of the optical lattice has been employed several times in the many-body regime, where instead of Rabi oscillations, one can measure frequency-dependent heating rates~\cite{Stoferle2004,Jordens2008}.

Specifically, we simulate the Hamiltonian $H=H_0+g\sin(\omega t)H_{\mathrm{lin}}(t)$ with 
\begin{equation}
  H_{\mathrm{lin}}(t) = \frac{1}{a_0}\sum_{\vec i}(i_x-N/2)c_{\vec i\sigma}\dagg c_{\vec i\sigma}.
  \label{eq:H_lin}
\end{equation}
In order to see clean Rabi oscillations, we should set $\omega$ close to the frequency of a dipole-allowed transition, say $\omega_{21}=\omega_{2p}-\omega_{1s}$ and drive weakly enough that transitions to other states are suppressed by the detuning to them.
To understand these conditions, note that the principal energy levels approximately have the energy~\cite{Zaslow1967}
\begin{equation}
  E_n = -\frac{\Ry}{(2n-1)^2} = -\frac{J}{a_0^2(2n-1)^2},
  \label{eq:En}
\end{equation}
where $\Ry=V_0^2/J=J/a_0^2$ is the Rydberg constant, $a_0$ is the Bohr radius in units of the lattice constant, and $J$ is the hopping strength. These energies are exact in the continuum $a_0\to\infty$, but the quality of approximation in finite systems and with finite $a_0$ depend on the chosen orbital. Nevertheless, they provide a good guide.
To get clean Rabi oscillations, the coupling must be weak.
For example, if we drive the transition from 1s to 2p, the detuning to the 1s--3p transition is $J(1/9-1/25)/a_0^2\approx0.07J/a_0^2$.
The matrix elements $\bra{2p}H_{\mathrm{lin}}\ket{1s}$ and $\bra{3p}H_{\mathrm{lin}}\ket{1s}$ depend somewhat on $a_0$ but are roughly of order 1.
Thus, we need $g\ll 0.1J/a_0^2$ to see Rabi oscillations.
Higher transitions are increasingly difficult to observe, because they require weaker coupling. 
For this reason we choose $a_0=1$ (small electron orbits), as this leads to shorter experiments.

We model spectroscopy numerically in the following way. 
We initialize the system in the ground state of hydrogen (one electron in the 1s orbital) and apply the oscillating Hamiltonian $H_0+g\sin(\omega t)H_{\mathrm{lin}}(t)$ (see \cref{eq:H_lin}) for a time $t$. 
We then detect the probability of the system to remain in the electronic ground state by first applying the inverse of the adiabatic preparation and then checking whether the atom returns to the initial lattice site.
This reduces the problem of detecting the ground state probability to measuring the probability of the atom to be localized on a specific site.
We repeat this process for different times from 0 to a total time $T$. 
Given sufficiently weak driving, the system will undergo oscillations if the drive frequency is tuned to an electronic transition, and otherwise remain mostly in the ground state. 
We then sweep the drive frequency and record the amplitudes of the observed Rabi amplitudes to detect the resonances of the system.
The result is shown in \cref{fig:spectroscopy}.

\section{Experimental implementation}
Our proposal requires fermionic atoms that can be trapped in an optical lattice, imaged with single-site resolution, Rydberg dressed, and that have at least two long-time coherent hyperfine levels (see illustration in \cref{fig:setup}).
These are all state-of-the-art capabilities, and several atomic species could serve this purpose.
Standard choices are the alkali atoms $^{40}$K or $^6$Li.
Alkaline-earth(like) atoms a number of favourable optical properties, such as ultra-narrow linewidth transitions that have enabled their use as clocks~\cite{Ludlow2015}, for quantum simulation~\cite{Mancini2015}, and make them promising candidates for quantum computing~\cite{Daley2008}.
Among the most common fermionic species are $^{87}$Sr, $^{171}$Yb, and $^{173}$Yb. 
Recently, coherent gates for individual (pairs of) $^{171}$Yb atoms have been achieved with high fidelity, exploiting the high coherence time of the nuclear spin~\cite{Ma2021}.
Other common species such as $^{87}$Sr~\cite{Daley2008,Barnes2021} may be equally suitable.
Proof-of-principle experiments of up to two electrons could also be done with bosons such as $^{87}$Rb.

The key figure of merit for experiments is the ratio of hopping rate divided by effective Rydberg decay rate, $M=J/\Gamma$.
It can be interpreted as the number of sites an atom can explore before decaying.
Preparing the ground states of larger molecules requires longer times or approaching the continuum limit by increasing $a_0$ requires longer simulation times.
To maximize $M$, it is favourable to use a light atomic species, such as $^6$Li, as done in Ref.~\cite{Guardado-Sanchez2021}. 

Rydberg decay can be suppressed by dressing with Rydberg states of higher principal quantum number $n$. The lifetime of low-angular momentum Rydberg states including black-body radiation scales as $n^3$. In contrast, the strength of the interactions scales as $n^4$. 
Thus, since we use Rydberg dressing to obtain an interaction strength comparable to hopping, we scale $\beta\propto n^{-1}$, such that $V_{\mathrm{int}}\propto n^4\beta^{-4}={}$const. As a result, the Rydberg decay rate relative to the hopping strength scales as $\Gamma\propto n^{-3}\beta^2\propto n^{-5}$. Using this scaling, we obtain that dressing to $n=70$ instead of $n=28$ as in Ref.~\cite{Guardado-Sanchez2021} would lead to a 100-fold improvement in $M$.

Another potential approach is stroboscopic dressing~\cite{ZeiherPhD}. If the dressing is turned on and off at a rate much larger than other relevant time scales, it is well approximated by continuous dressing, but with the values for $\beta^2$ and $\beta^4$ replaced by their time averaged values.
If we assume for simplicity that dressing is on for a fraction $\eta$ of the time and otherwise off, we find that to keep the same effective interaction strength, we need to increase $\beta\to\beta_\eta = \beta \eta^{-1/4}$.
However, the induced Rydberg decay will then be reduced $\Gamma\to\Gamma_\eta=\Gamma\eta^{1/2}$, such that $M\propto\eta^{1/2}$.

Finally, it would be worth exploring how weak the optical lattice can be before problematic effects appear.
For example, the standard condition is that next-to-nearest neighbour hopping must be negligible, but this is perhaps not strictly necessary here, especially since in real atoms and molecules there is a relativistic correction to the kinetic energy $\sim p^4$, which would be generated through next-to-nearest interactions.
In principle, the scheme presented here could in principle even work in the continuum, although it is not clear how exactly one would control the kinetic energy term and implement the nuclear potential.

\section{Conclusion}
We have shown that Rydberg dressing in standard ultracold lattice atom experiments allows for the simulation of pseudo-quantum chemistry, which differs from real quantum chemistry by being two-dimensional and having a different scaling law for the Coulomb interaction.
Nevertheless, pseudo-quantum chemistry inherits key phenomenology from quantum chemistry, including (anti-)bonding orbitals and the potential for spectroscopy.
The key experimental figure of metric that determines how well such simulations can be done is the ratio of Rydberg decay rate to hopping strength.
We show using parameters from existing experiments that already current experiments could prepare the ground state of small molecules.
Since it is realistic for the figure of merit to increase substantially in the future, 
this motivates using this platform to explore quantum chemistry physics experimentally, to gain insights into complex electronic processes, and to benchmark classical codes.

Future work should explore how spin-orbit interactions, or other important corrections (e.g., relativistic corrections) could be added to the simulator. Moreover, the adiabatic paths used in our numerical simulation can likely be improved, especially when tailored to given specific experimental capabilities.
We have also hardly explored $\alpha=3$, which may feasibly be obtained through Rydberg dressing, and could be advantageous as the resulting interactions is stronger, which could result in a larger ratio of hopping to decay rate.
It would also be interesting to go more into detail about the physics that differs from normal quantum chemistry. 
An obvious example is that neither the nuclear potential nor the electronic repulsion obey Gauss' law, which means that in principle arbitrarily many electrons could fit around a nucleus.

Going beyond the analogy with quantum chemistry, it would be interesting to understand how the physics of this platform is modified in the presence of additional interactions, disorder, or unavoidable dissipation.
Finally, the few-body, low-density regime of ultracold atom experiments may yield other fruitful research directions.
For example, when the repulsive interaction dominates the kinetic energy term and the atoms are sufficiently dense, the ground-state configuration ought to be related to packing of circles in two dimensions.
These configurations can also be obtained adiabatically. 

\begin{acknowledgments}
	We thank Pascal Weckesser and Simon Hollerith for helpful discussions about experimental considerations.
	We acknowledge funding from ERC Advanced Grant QUENOCOBA under the EU Horizon 2020 program (Grant Agreement No. 742102).
	This research is part of the Munich Quantum Valley, which is supported by the Bavarian state government with funds from the Hightech Agenda Bayern Plus.
\end{acknowledgments}

\bibliography{library}{}
\end{document}